\documentclass[useAMS,usenatbib,onecolumn]{mn2e}

\usepackage{amsmath}
\usepackage{comment}
\usepackage{graphicx}
\usepackage{rotating}
\usepackage{color}
\usepackage{aas_macros}
\usepackage{amsfonts}
\newcommand{\beq}{\begin{equation}}
\newcommand{\enq}{\end{equation}}
\newcommand{\beqa}{\begin{eqnarray}}
\newcommand{\enqa}{\end{eqnarray}}
\newcommand{\beit}{\begin{itemize}}
\newcommand{\enit}{\end{itemize}}
\newcommand{\bem}{\begin{pmatrix}}
\newcommand{\enm}{\end{pmatrix}}

\newcommand{\vecx}{\mathbf{x} }

\newcommand{\veck}{\mathbf{k} }

\newcommand{\Tr}{\mathrm{Tr}}

\newcommand{\lat}{\left\langle}
\newcommand{\rat}{\right\rangle}
\newcommand{\av}[1]{\lat #1 \rat}

\newcommand{\obs}{\textrm{obs}}

\newcommand{\true}{\textrm{true}}
\newcommand{\cst}{\textrm{const}}

\newcommand{\lb}{\left [}
\newcommand{\rb}{\right ]}
\newcommand{\lp}{\left (}
\newcommand{\rp}{\right )}

\renewcommand{\max}{\mathrm{max}}
\renewcommand{\min}{\mathrm{min}}

\newcommand{\bes}{\begin{sideways}}
\newcommand{\ees}{\end{sideways}}

\renewcommand{\bbeta}{\boldsymbol {\eta}}
\renewcommand{\bmu}{\boldsymbol {\mu}}

\newcommand{\vecn}{\textbf{n}}

\newcommand{\vecun}{\mathbf 1}
\newcommand{\Nbin}{N_{\textrm{bin}}}

\newcommand{\kappatrue}{ \kappa}
\newcommand{\kappaobs}{\kappa^{\textrm{obs}}}
\newcommand{\vecell}{\boldsymbol \ell}
\newcommand{\vecm}{\textbf m}

\title[Super-surveys modes and cosmic shear]{The impact of super-survey modes on cosmological constraints from cosmic shear fields}

\author[Carron, Szapudi]{J. Carron\thanks{E-mail:
carron@ifa.hawaii.edu}, I. Szapudi  \\
Institute for Astronomy, University of Hawaii, 2680 Woodlawn Drive, Honolulu, HI, 96822}
\begin{document}

\date{\today}

\pagerange{\pageref{firstpage}--\pageref{lastpage}} \pubyear{2014}

\maketitle

\label{firstpage}

\begin{abstract}
Owing to the mass-sheet degeneracy, cosmic shear maps do not probe directly the Fourier modes of the underlying mass distribution on scales comparable to the survey size and larger.
To assess the corresponding effect on attainable cosmological parameter constraints,
we quantify the information on super-survey modes in a lognormal model and, when interpreted as nuisance parameters, their degeneracies to cosmological parameters. Our analytical and numerical calculations clarify the central role of super-sample covariance (SSC) in shaping the statistical power of cosmological observables. 
Reconstructing the background modes from their non-Gaussian statistical dependence to small scales modes yields the renormalized convergence. This  diagonalizes the spectrum covariance matrix, and the information content of the corresponding power spectrum is increased by a factor of two over standard methods. Unfortunately, careful calculation of the Cramer-Rao bound shows that the information recovery  can never be made complete, any observable built from shear fields, including optimal sufficient statistics, are subject to severe information loss, typically $80\%$ to $90\%$  below $\ell \sim 3000$ for generic cosmological parameters. The lost information can only be recovered from additional, non-shear based data. Our predictions hold just as well for a tomographic analysis, and/or full sky surveys.
\end{abstract}

\section{Introduction}
During the nonlinear evolution from Gaussian initial conditions, Fourier modes of the matter density field gradually develop statistical dependence, even though still uncorrelated due to statistical homogeneity. Thus a Gaussian approximation to estimating the cosmic variance and covariance of key cosmological statistics such as the power spectrum no longer holds.  Yet, a precise understanding of cosmic (co)variance is necessary for the ultimate success of the ambitious upcoming wide-field surveys targeting cosmic acceleration such as LSST\footnote{http://www.lsst.org/lsst}, Euclid\footnote{http://sci.esa.int/euclid}, and WFIRST\footnote{http://wfirst.gsfc.nasa.gov}. Perhaps the most intriguing non-Gaussian signature of the dark matter field, the principal aim of this paper, is the large impact of super-surveys modes, corresponding to scales comparable to and larger than the survey size, on the cosmological constraints from the observed smaller scales modes. Early work \citep[e.g.,][]{Peebles1980}, has shown that on large scales the correlation function is affected by the local estimate of the average density, and several estimators for the correlation function \citep{Peebles1980,Hamilton1992,LandySzalay1993, SzapudiSzalay1998} were constructed to be less sensitive to the ``integral constraint'' characterized by the variance of the number density of the survey proportional to the average of the two-point correlation function over the survey volume \citep[e.g.,][]{SzapudiColombi1996}.
 \cite{RimesHamilton2005, RimesHamilton2006,NeyrinckEtal06} identified the resulting loss of information, the  ``information plateau'', in the context of Fisher information theory,  grossly contradicting Gaussian expectations.
More recently, the impact of super-surveys modes on the power spectrum covariance (the super-sample covariance, hereafter SSC) was elegantly characterized by the response of the the power spectrum to a change in the background density $\delta_b$ by \cite{TakadaHu2013,LiEtal2014}.
\newline
\indent
Both galaxy clustering and cosmic shear \citep[see][for reviews]{BartelmannSchneider01,MunshiEtal2008,WeinbergEtal2013}  probe the dark matter distribution in the Universe. However, there is a clear distinction between the two in the context of SSC. As pointed out most clearly by \cite{DePutterEtal2012}, the coherent fluctuations of the galaxy density field and the total number number of objects  (a direct probe of $\delta_b$) reduce drastically the  response of the spectrum and therefore moderate the impact of the SSC. In contrast, cosmic shear fields do not feel the mass background mode $\delta_b$ due to the mass sheet degeneracy. Thus there is no analog of observed mean density for cosmic shear, and one cannot simply recalibrate the power spectrum or other observables to mitigate the effects of super-survey fluctuations, and the corresponding SSC. The ultimate result, as shown later, is severe information loss. A promising approach to mitigate the impact of SSC in weak-lensing survey is using additional data sensitive to the background modes in a joint analysis. An approach using cluster counts has been advocated by \cite{TakadaSpergel2014}. In this work we explore the limits when using shear maps only. 
\newline
\indent
The principal aim of this paper is to quantify the impact of the absence of $\delta_b$ on the total information content of shear fields, and to explore
the possibility of self-calibration to at least partially recover the lost information due to SSC.
The key idea is the following: cosmic shear measures the Fourier modes of the convergence field $\kappa$ with the exception of the zero-mode $\delta_b$. Using the non-Gaussian statistical dependence of the Fourier modes one can partially reconstruct the zero-mode, thus alleviating some of the information loss due to SSC. \newline
\indent
We proceed with analytical methods, using a lognormal model \citep{ColesJones1991} for the convergence field statistics. The field $\ln(1 + \kappa/|\kappa_0|)$, where $|\kappa_0|$ is the minimal value of the convergence, is modeled as a Gaussian field with the appropriate two-point function and ensemble mean. Low dimensional lognormal probability density functions (PDFs) have been known to reproduce accurately the dark matter and convergence PDFs \citep[e.g.]{BernardeauKofman1995, TaruyaEtal2002}. Most importantly, the lognormal model has been shown explicitly to reproduce well the impact of SSC on the spectrum information content \citep{TakahashiEtal2014, CarronEtal2014}. Finally, our simple model will allow us to obtain analytical insights into the effect of background modes. In particular, the joint Fisher matrix describing the degeneracy of $\delta_b$, considered as a nuisance parameter, to cosmological parameters is obtained exactly in this paper, allowing us understanding of the degradation in the total information content of the shear field beyond the power spectrum.
\newline
\indent
The paper is organized as follows. Section \ref{section2} presents keys formulae on how well background modes can be reconstructed from small scale modes in lognormal fields, together with the degeneracy of these modes with model parameters. Section \ref{Sim} presents implications and tests of these results. After discussing our simulations of lognormal fields we implement in section \ref{simsim} the background mode reconstruction yielding the renormalized convergence (RC). We discuss in sections \ref{covariance} and \ref{informationspectrum} the covariance matrices and information of the RC spectrum. Section \ref{totalinfo} contains the main result of this paper: the impact of the absence of the background mode on the total information content of the field. Section \ref{fullsky} deals with the case of full sky tomographic surveys showing that the results are essentially unchanged. Finally, we summarize and conclude in section \ref{conclusion}. The appendix presents additional technical details on the derivation of our results.
\section{Reconstructing the background modes}\label{section2} 
Let us consider a survey volume $V$ regularly sampled at $d$  number of points, $x_i$. Thus there are exactly $d$ discrete Fourier modes. A field $\phi$ taking values on the grid can be written
\beq \label{FT}
\phi(x_i) = \frac 1 V\sum_\veck \tilde \phi_\veck\: e^{i\veck\cdot x_i}\quad  \textrm{ with } \quad\tilde \phi_\veck = \frac{V}{d} \sum_{x_i} \phi(x_i) e^{-i\veck\cdot x_i}. 
\enq
Our discrete Fourier convention is such that in the asymptotic regime $d, V \rightarrow \infty$ we would recover
\beq
\phi(x) \rightarrow \int \frac{d^n k}{\lp 2\pi\rp^n} \tilde \phi(\veck) \: e^{i\veck\cdot \vecx}\quad \textrm{ with } \quad \tilde \phi(\veck) = \int d^nx\: \phi(x) e^{-i\veck \cdot x}.
\enq
The zeroth mode $\tilde \phi_0$ of the volume encapsulates the background super-survey modes, we write for it $\phi_b$. We also define the useful dimensionless spatial average
\beq
\bar \phi = \frac{ \phi_b}{V} = \frac 1 d \sum_{x_i}\phi(x_i).
\enq
\subsection{Generalities}
Assuming a specific form for the $d$-variate PDF $p_\delta$ describing the joint occurrences of the fluctuation field values, we can ask how well we expect to be able to reconstruct an unseen zero mode $\tilde\delta_0$, and how it correlates with other parameters. We first note that the likelihood for the zero mode given the observation of the non-zero modes is given by the $p_\delta$ itself
\beq
\ln p(\tilde \delta_0 | \tilde  \delta_{\veck \ne 0}) \propto \ln p_\delta(\delta).
\enq
The curvature of left hand side as a function of $\tilde \delta_0$ is the inverse variance of the posterior for the background mode. Using most conveniently the dimensionless local average density $\bar \delta =  \tilde \delta_0 / V$, this curvature is on average nothing else than the Fisher information content $-\av{\partial^2_{\alpha^2} \ln p}$ of the field on this parameter. Since $\delta(x_i)  = \bar \delta + \sum_{\veck \ne 0} \tilde \delta_\veck\:e^{i\veck\cdot x_i}$ follows $\partial \delta(x_i)/\partial {\bar \delta} = 1$ and thus
\beq
\frac{\partial \ln p_\delta}{\partial \bar \delta} = \sum_{i = 1}^d \frac{\partial \ln p_\delta}{\partial \delta_i} \quad \textrm{and} \quad \frac{\partial^2 \ln p_\delta}{\partial \bar \delta^2} = \sum_{i,j = 1}^d \frac{\partial^2 \ln p_\delta}{\partial \delta_i\partial  \delta_j} 
\enq
Therefore in full generality holds
\beq \label{Fdeltab}
F_{\bar \delta \bar \delta} = - \av{ \lp \sum_{i,j = 1}^d \frac{\partial^2}{\partial \delta_i\partial \delta_j}\rp  \ln p_\delta}.
\enq
Likewise the degeneracy to a model parameter is described by
\beq \label{Fdeltabparam}
F_{\alpha \bar \delta } = - \av{ \lp \sum_{i = 1}^d \frac{\partial^2}{\partial \delta_i\partial \alpha}\rp  \ln p_\delta}.
\enq
\subsection{Background modes in lognormal fields}
\label{Modesreconstruction}
We now state the results for the background mode extraction in lognormal fields upon which a fair amount of our subsequent considerations is based. For full generality, we consider a number $\Nbin$ of jointly lognormal fields $\delta_n(x_i)$ with zero ensemble mean (in a tomographic analysis of a weak-lensing survey these fields represent the rescaled convergence fields $\delta_n(x_i) = \kappa_n(x_i)/|\kappa_{0,n}|$ where $\kappa_{0,n}$ is the total weight of the $n$th lensing kernel, see section \ref{Sim} later for definitions and more details). The fields are described statistically by $\Nbin \times \Nbin $ hermitian spectral matrices $P^{nm}_\delta(\veck)$
\beq
\av{\tilde \delta_n(\veck) \tilde \delta^*_m(\veck')}  =V \delta_{\veck\veck'} P^{nm}_\delta(\veck)
\enq
Assuming periodic boundary conditions, these are Fourier transforms of the $d \times d $ covariance matrices $\xi_\delta^{nm}$. 
The log-densities $A_n(x_i) = \ln (1 + \delta_n(x_i))$ form jointly Gaussian fields with covariance matrices $\xi_A^{nm}(x_i-x_j) = \ln (1 + \xi_\delta^{nm}(x_i-x_j))$, and ensemble mean vector $\av{A_n} = -\sigma^2_{A,n}/2$. The Fourier transforms of $\xi_A$ are the spectral matrices $P_A(\veck)$, which are the most natural variables to describe the lognormal field. There are $\Nbin$ background modes $\bar \delta_n$.
\newline
Appendix \ref{AppendixInfo} details the derivation of the following results. The Fisher information matrix on the background modes is
\beq \label{Fdb}
\begin{split}
F_{\bar\delta_n\bar \delta_m} &\: =   e^{\sigma^2_{A_n}}e^{\sigma^2_{A_m}} \lb \delta_{nm} \:d \:e^{\sigma^2_{A_n}} + \sum_\veck \lb P^{-1}_{A}(\veck) \rb_{nm} P^{nm}_{\delta}(\veck) +  \lp \frac  {P_A(0)}V \rp^{-1}_{nm} \rb .  
\end{split}
\enq
The last term is the Gaussian limit where the modes are statistically independent. It only depends on the survey geometry. The first terms are due to mode coupling and are directly proportional to the number $d$ of modes available. For a given survey volume the zeroth mode can thus be reconstructed meaningfully from the others provided the resolution $k_\max$ is large enough for the first terms to be dominant. This happens roughly at the critical value $\bar n P_A(0) \sim 1$, where $\bar n = d/V$ is the number density of grid points, in one to one correspondence to $k_\max$.
The degeneracy to a model parameter is described by the Fisher matrix element
\beq \label{Fadb}
\begin{split}
F_{\bar\delta_n \alpha} &\:= e^{\sigma^2_{A,n}}\lb \sum_{\veck} \lp P_A^{-1}(\veck)\frac{P_A(\veck)}{\partial \alpha} \rp_{nn} + \sum_{m  = 1}^{\Nbin} \lp \frac {P_A(0)} V \rp^{-1}_{nm} \lp \frac 12\frac{\partial \sigma^2_{A,m}}{\partial \alpha} \rp \rb.
\end{split}
\enq
In stark contrast to a Gaussian field (for which $F_{\bar \delta \alpha} = 0$, see \eqref{eqA2}), we find that all modes correlate equally with $\bar\delta$, and $F_{\alpha \bar \delta}$ is typically proportional to the number of modes present.  Finally, the information matrix for model parameters in the lognormal field is identical to the well known formula for Gaussian fields
\beq \label{Fln}
\begin{split}
F_{\alpha \beta} &\:=  \frac 12 \sum_\veck \Tr \lb P_A^{-1}(\veck) \frac{\partial P_A(\veck)}{\partial \alpha}P_A^{-1}(\veck)\frac{\partial P_A(\veck)}{\partial \beta} \rb +  \lp \frac 12 \frac{\partial \sigma^2_A}  {\partial \alpha}\rp^t \lp \frac{P_A(0)}{V} \rp^{-1}\lp \frac 12 \frac{\partial \sigma^2_A}  {\partial \beta}\rp
\end{split}
\enq
This last equation follows directly from the fact that the log-density fields $A$ are jointly Gaussian with mean vector $-\frac 12\sigma_A^2$. 
Before proceeding, a word of caution. Eqs. \eqref{Fadb} and \eqref{Fln} hold assuming that $\kappa_0$ is independent of the parameters of interest. This holds throughout this paper as we will be interested in $\sigma_8$ and the spectral index $n_s$, $\kappa_0$ being purely geometric. The appendix provides the complete expressions including the derivatives of $\kappa_0$ for completeness.
\section{Tests and implications} \label{Sim}
We test the background mode reconstruction and the above results in a large ensemble of simulations of lognormal fields. We simulate a square survey of $L = 10$ degrees on the side with $d = N^2$ points on the flat sky. To generate each map as realistically as possible we use the circulant embedding method \cite[for details see][]{CarronEtal2014}. The spectrum assigned to the discrete Fourier modes is defined through the following equation 
\beq 
\begin{split} \label{Pamap}
P^{\textrm{discrete}}_{\delta,A}(\vecell) = \frac{V}{d} \sum_{x_i} \xi_{\delta,A}(x_i) e^{-i\vecell\cdot x_i} \quad \textrm{    with grid points  }
\quad x_i & = \frac L N  \bem i \\ j \enm,\quad i,j = -\frac N 2,\cdots,\frac N 2-1 
\end{split}
\enq
The circulant embedding method uses Fast Fourier Transform (FFT) algorithms and is thus very fast. In contrast to more usual FFT-based methods that sample directly the power spectrum it is also accurate : the field values on any subgrid of half the side follows the exact (non-periodic) target covariance matrices. In particular the super-surveys modes are correctly accounted for, even though, for simplicity, in this paper we use the full periodic box  for which Eqs. \eqref{Fdb}, \eqref{Fadb} and \eqref{Fln} hold exactly, with $P_A,P_\delta$ given by Eq. \eqref{Pamap}. The fluctuations of the density and log-density in the volume are described by $P^{\textrm{discrete}}_{\delta,A}(0)$, which according to Eq. \eqref{Pamap} get contributions from the continuous spectra at scales comparable to the survey volume. In the following we suppress the superscript of the discrete spectra as it should be clear from the context what spectrum is meant. The number of modes in our simulations is $d = 256^2$. The periodicity of the simulated volume is clearly motivated by convenience and is unphysical but the continuation of our results for a full sky survey will be easy to establish. The two-point function of the convergence required in Eq. \eqref{Pamap} as well as $\kappa_0$ is set to match that of a vanilla flat $\Lambda$CDM Universe. It is calculated Legendre transforming  the Limber approximation to the projected power spectrum prediction  \citep{Kaiser92,BartelmannSchneider01}
\beq \label{3dspec}
P_{\kappa}(\ell) =  \int_{0}^{\chi_s}d\chi \frac{g^2(\chi)}{\chi^2}P^{3d}_\delta\lp  \frac{\ell} {\chi}, \chi\rp,\quad g(\chi) = \frac 32 \Omega_m\lp \frac{H_0}{c}\rp^2\:\frac{\chi}{a(\chi)} \lp1-\frac \chi \chi_s \rp, \quad \kappa = \int_0^{\chi_s} d\chi \:g(\chi)\delta^{3d}(\chi),
\enq
where $\chi$ is the comoving distance. These approximations are
appropriate for the multipole range of the map  ($\ell_\min = 36$ and $\ell_\max = 6517$). We use a single source redshift $z_s = 1$ for the simulated maps. The two-point function is filtered with a spherical top-hat filter of radius $\ell_{\textrm{cell}} / \sqrt \pi$ to account for the pixelisation. The 3d matter power spectrum in Eq. \eqref{3dspec} is generated using the revised version of the halo-fit model \citep{SmithEtal03,TakahashiEtal12} as implemented in the CAMB\footnote{http://camb.info/}\citep{LewisEtal2000} software. Finally, the minimal value of the convergence $\kappa_0$ is given by
\beq
\kappa_0  = - \int_0^{\chi_s} d\chi\: g(\chi),
\enq
making $\delta = \kappa/|\kappa_0|$ a true averaged $\delta^{3d}$ with weight function normalized to unity.
A simulation of an observed convergence field $\kappaobs$ (the $\vecell \ne 0$ modes of the convergence field) is simply performed as follows:
i) we generate a d-dimensional lognormal vector $x_i = 1 + \delta_i$ following the algorithm of \cite{CarronEtal2014}  ii) we transform it to $\kappa = |\kappa_0|\delta $ and finally iii) the background mode is set to zero in all maps according to
$ \kappaobs=  \kappatrue - \bar \kappa.$
Some key parameters for the simulated maps are $\kappa_0 = -0.06, \sigma^2_A = 0.06,  P_A(0)/ V \sim 3.6\cdot10^{-4} $
\subsection{Reconstruction of the background mode}\label{simsim}
\begin{figure}
\begin{center}
\includegraphics[width = 0.45\textwidth]{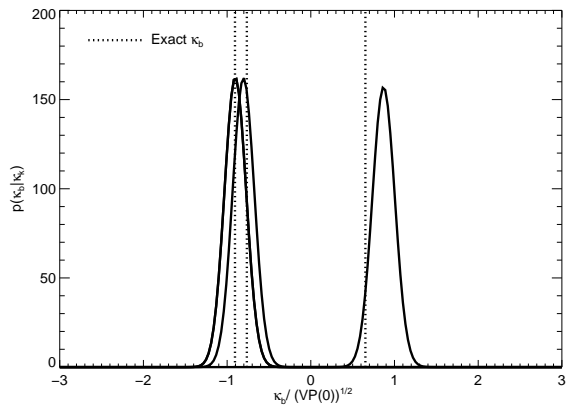}
\caption{\label{3PDFs} The posterior probability for the unobserved background mode $\kappa_b$ given the observed Fourier modes of the convergence, for three independent simulations of lognormal convergence fields, for a $\Lambda $CDM convergence power spectrum in a $10 \times 10$ deg$^2$ survey with source redshifts $z_s = 1$. The vertical dotted lines indicates the true value of the $\kappa_b$ in the map. The horizontal axis is in units of the root of  the unconstrained variance $P_\kappa(0)/V$ of $\kappa_b$. The variance of each PDF follows very closely the analytical prediction given by Eq. \eqref{Fdb}.}
\end{center}
\end{figure}

We now test the idea of reconstructing $\bar \kappa$.
The likelihood for the background mode given the non-zero modes of the volume is given by their joint PDF. In the lognormal model this is
\beq
\begin{split}
\ln p(\bar \kappa | \kappaobs)& =  -\frac12 \sum_{i,j = 1}^d \lp A_i(\bar \kappa) + \sigma^2_A / 2 \rp \lb  \xi_{A}^{-1} \rb_{ij} \lp   A_j(\bar \kappa)+ \sigma^2_A / 2 \rp  - \sum_{i = 1}^d A_i(\bar \kappa) + \cst.
\end{split}
\enq
In this equation the log-density field $A$ is to be seen as a function of the background mode according to
\beq
A_i(\bar \kappa) = \ln \lp 1 +\frac{ \kappaobs_i + \bar\kappa}{|\kappa_0|} \rp.
\enq
The likelihood can be evaluated most simply using the periodicity of the box. Following our discrete Fourier transform conventions it becomes
\beq
\label{lnp}
\begin{split}
\ln p(\bar \kappa | \kappaobs)& =  - \frac12 \sum_{\vecell} \frac{ |\tilde A_{\vecell}(\bar \kappa) +\frac 12 \sigma^2_A V\delta_{\vecell 0} |^2}{VP_A(\vecell)}  - \frac d V \tilde {A_0}(\kappa_b) + \cst.
\end{split}
\enq
Fig. \ref{3PDFs} shows the posterior PDF for three different simulated maps. They are almost perfect Gaussians, with variance matching the Fisher prediction of Eq. \eqref{Fdb} within 2-3\% for each simulation. The vertical lines show the exact background mode of the corresponding maps.  The horizontal axis is in units of $\lp VP_\delta(0) \rp^{1/2}$, the unconstrained root variance of $\kappa_b = \bar \kappa V$.
From the observed maps we define the estimator of the background mode $\hat {\bar \kappa}$ as the argument of the maximum of $\ln p$,
\beq
\label{defkb}
\hat {\bar \kappa } =\underset{{\bar \kappa}}{\text{arg max}} \ln p(\bar \kappa | \kappa^\obs)  
\enq
We locate for each simulated map the maximum using a standard Newton-Rapshon non-linear solver scheme. To that aim Eq. \eqref{lnp} and its derivatives can all be efficiently performed using FFT algorithms. We found the resulting non-linear equation to be very well behaved, iterations with starting point $0$ converging to sufficient accuracy after only a few steps.
\begin{figure*}
\begin{center}
\includegraphics[width = 0.35\textwidth]{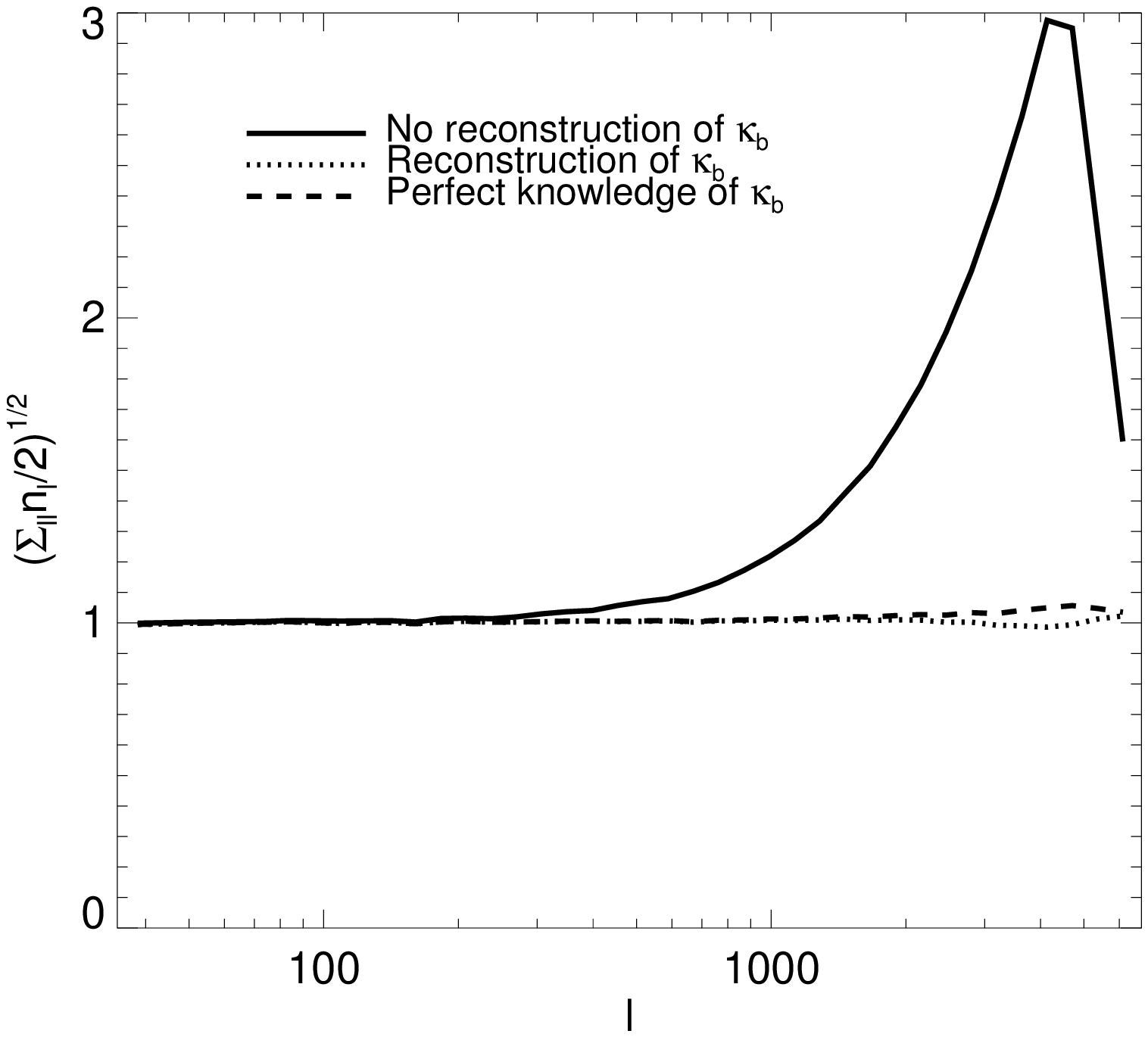}
\includegraphics[width = 0.3\textwidth]{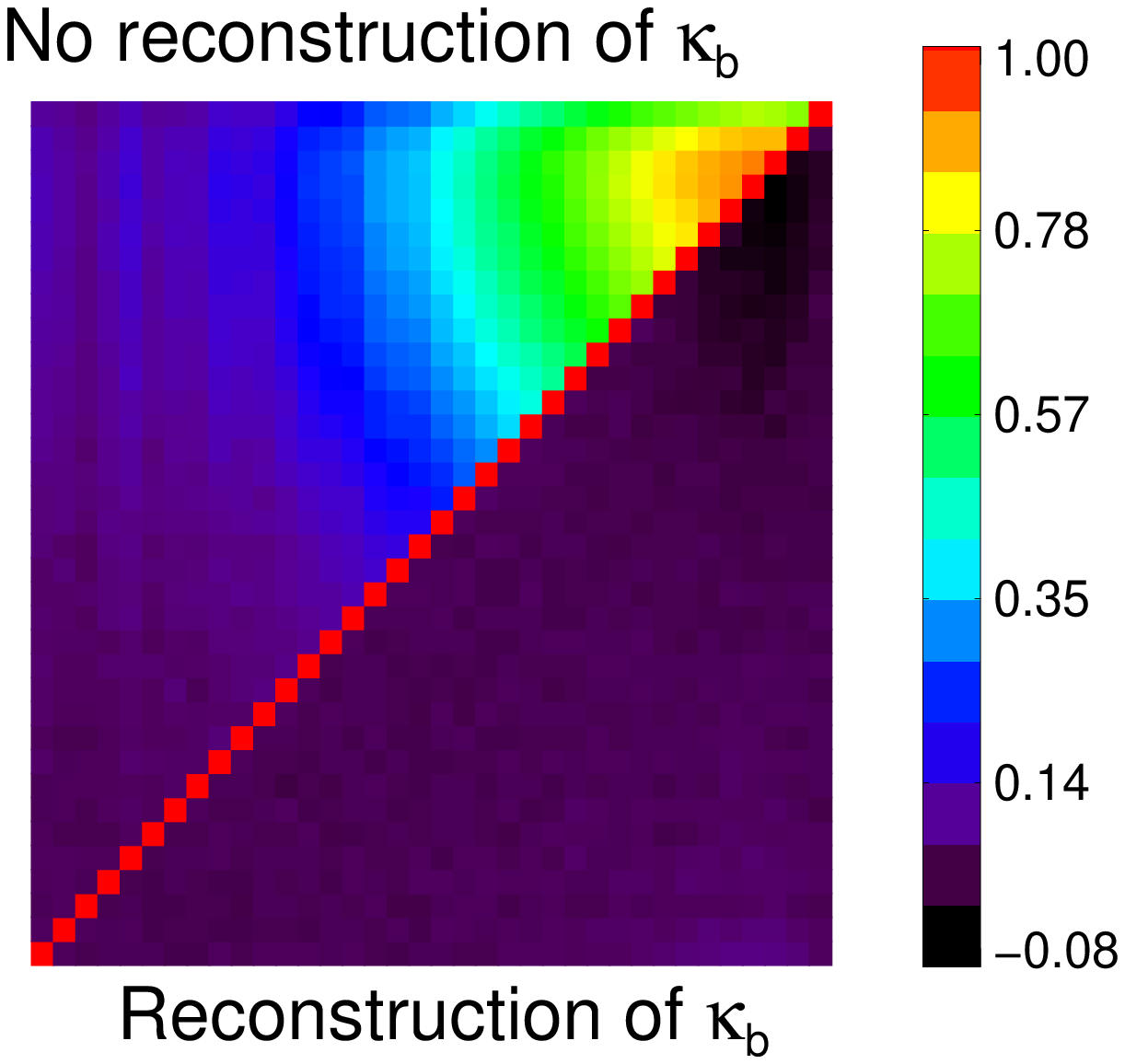}
\includegraphics[width = 0.3\textwidth]{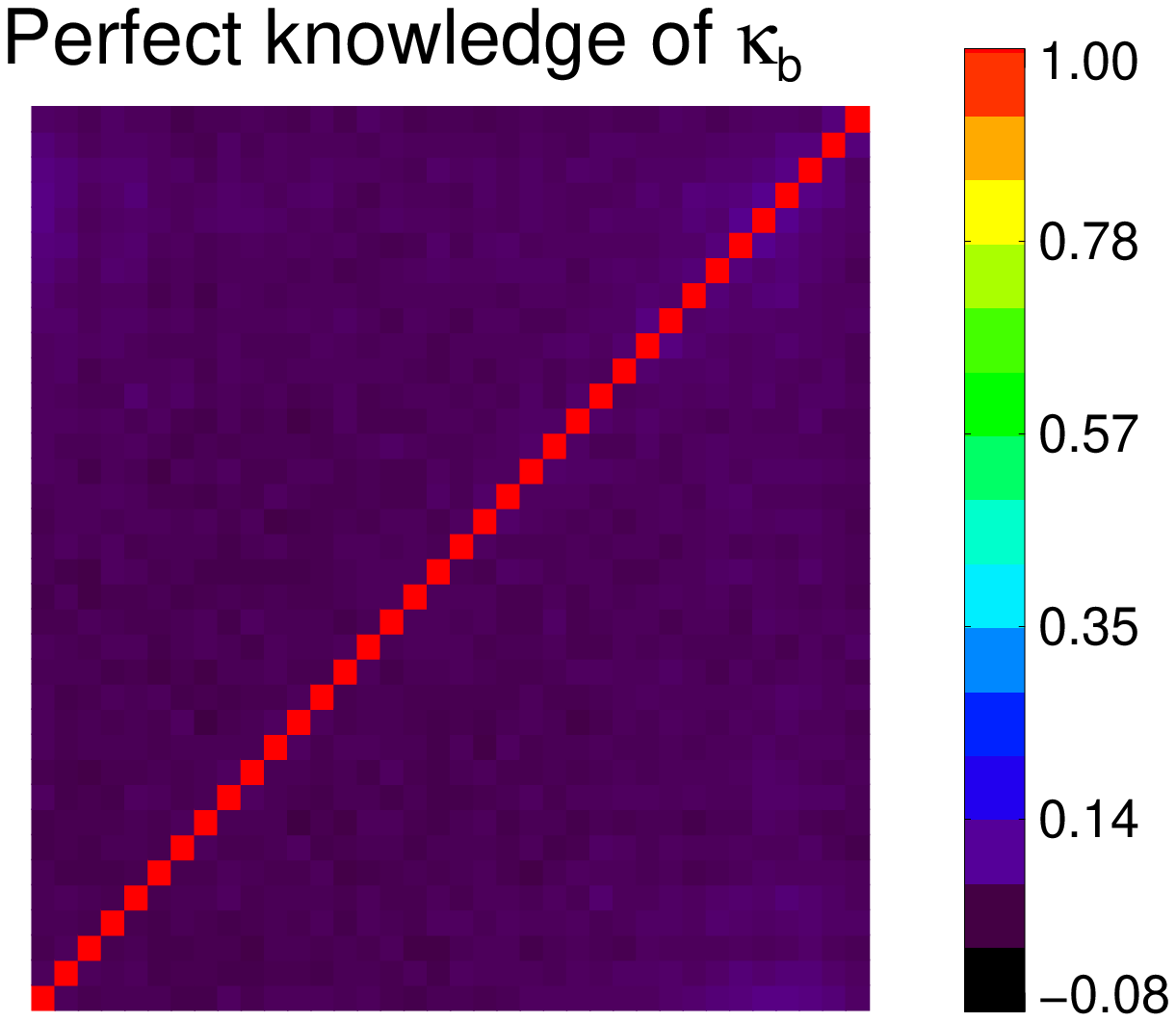}
\caption{Leftmost panel : the diagonal elements $\sqrt{ \Sigma_{\ell\ell} N_\ell/2}$ of the covariance matrix of a $\Lambda $CDM convergence power spectrum extracted from a $10 \times 10$ deg$^2$ volume with source redshifts $z_s = 1$ in three different cases, assuming lognormal field statistics. Unity corresponds to the Gaussian regime. In case i) no attempt to correct for the unseen background Fourier made is is made (solid), and the (co)variance is dominated by super-sample covariance. In case ii) the background mode is reconstructed and used to recalibrate the spectrum (dotted). Case iii) (dashed) shows the case of a perfectly known background mode for comparison. The center panel shows the matrix of correlation coefficient $\Sigma_{\ell \ell'}/\sqrt{\Sigma_{\ell \ell}\Sigma_{\ell' \ell'}}$ of the covariance matrix in cases i) (upper left corner) and ii) (lower right corner). The reconstruction of the background mode diminishes the SSC induced correlations to the level of case iii) in the rightmost panel. The color code is the same on both right-hand panels. Note that shape noise is not considered.\label{CovCorr}}.

\end{center}
\end{figure*}
\subsection{Spectrum and covariance matrix} \label{covariance}
It is natural to use knowledge of the reconstructed background mode to recalibrate the spectrum, in analogy to dividing with the average density for large scale structure maps. For a generic map $\phi$, we estimate its power spectrum as follows
\beq
\hat P^\phi(\ell) = \frac{1}{V}\frac{1}{N_\ell}\sum_{\vecell' \in \Delta(\vecell)} \left | \tilde \phi(\vecell')\right|^2.
\enq
The quantity $N_\ell$ is the number of modes in the  bin $\Delta(\ell)$. We use 40 bins equally spaced in $\ln \ell$ over the full range probed by the map. We compare the statistical properties of three slightly different maps of the convergence defined as 
\beq
\phi(x_i) = \frac{\kappaobs(x_i)}{ 1 + \hat {\bar \delta}} \quad  \textrm{ with } \quad
\textrm{case i)  }\hat{ \bar\delta} = 0,\quad\textrm{case ii) (RC),  }\hat{ \bar\delta} =  \frac{\hat { \bar\kappa}}{|\kappa_0|},\textrm{  and case iii)  } \hat{ \bar\delta} = \frac{\bar \kappa}{|\kappa_0|}
\enq
Case i) introduces no correction to account for SSC, which is the standard approach. Case ii) corrects with the background mode reconstructed according to our estimator defined in Eq. \eqref{defkb}.  The resulting spectrum is the renormalized convergence (RC) spectrum. Case iii) is the hypothetical case of a perfectly known background mode for comparison. We obtain expectation values $\av{\hat P(\ell)}$ and covariance matrices $\Sigma_{\ell \ell'} = \av{\hat P(\ell) \hat P(\ell')} - \av{\hat P(\ell)}\av{\hat P(\ell')} $ using a large number $(> 10^3)$ of simulations. The covariance matrices are shown on Fig. \ref{CovCorr}. The leftmost panel shows the diagonal elements $\sqrt{ \Sigma_{\ell\ell} N_\ell/2}$ as the solid, dotted and dashed lines for case i), ii), and iii). The prefactor of the diagonal is such that results unity for a Gaussian map. The center panel shows the correlation coefficient matrix of the covariance $\Sigma_{\ell \ell'}\sqrt{\Sigma_{\ell\ell} \Sigma_{\ell' \ell'}}$ of the spectrum in the standard approach (upper left corner) and for the RC spectrum (lower right corner). Finally, the rightmost panel shows the correlation coefficient matrix in case iii). It is obvious from these figures that the reconstruction works very well at decorrelating the modes, recovering a covariance matrix closer to case iii), itself not very far from the Gaussian case.
\subsection{Spectrum information content}\label{informationspectrum}
\begin{figure}
\begin{center}
\includegraphics[width = 0.45\textwidth]{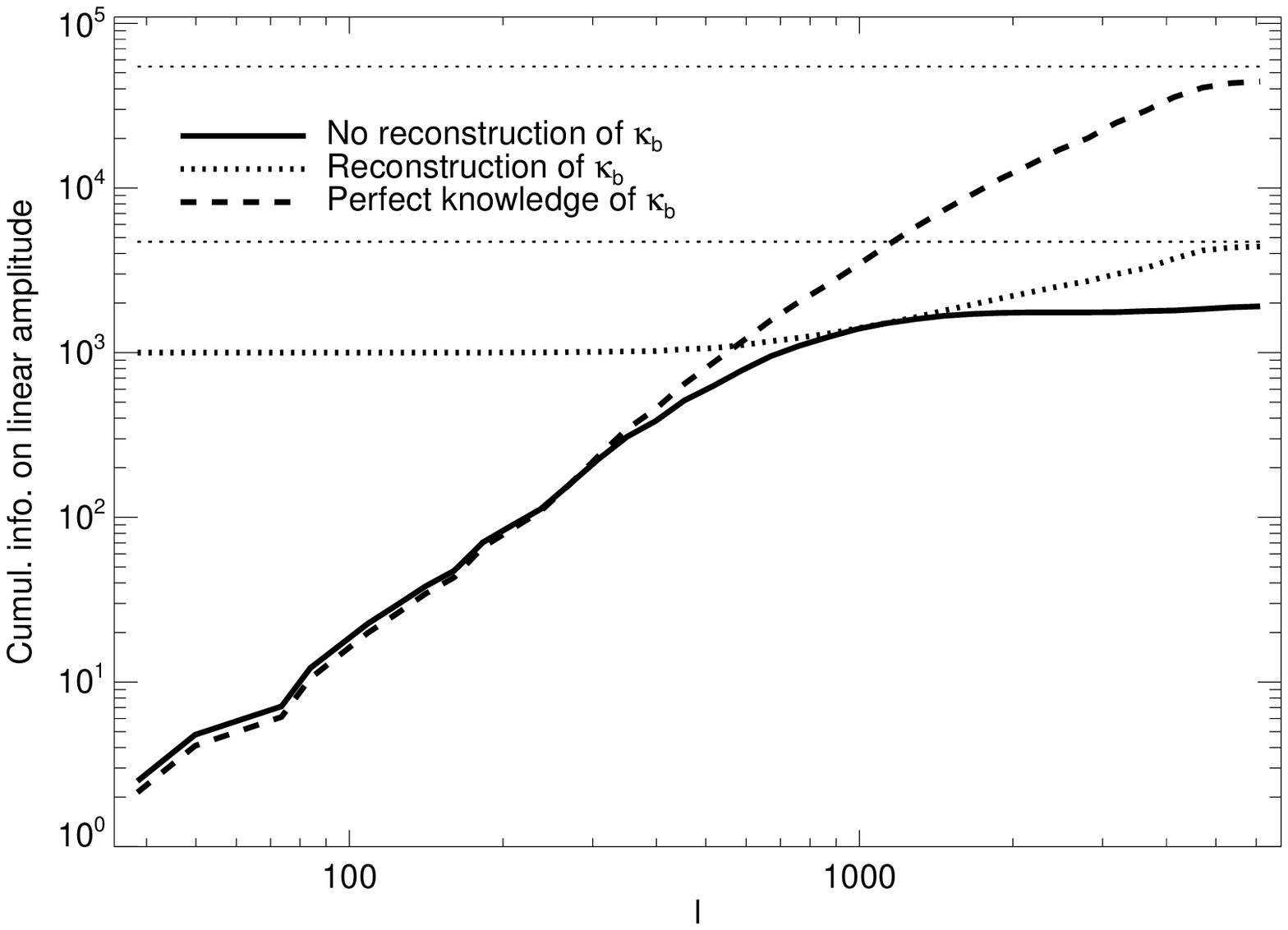}
\includegraphics[width = 0.45\textwidth]{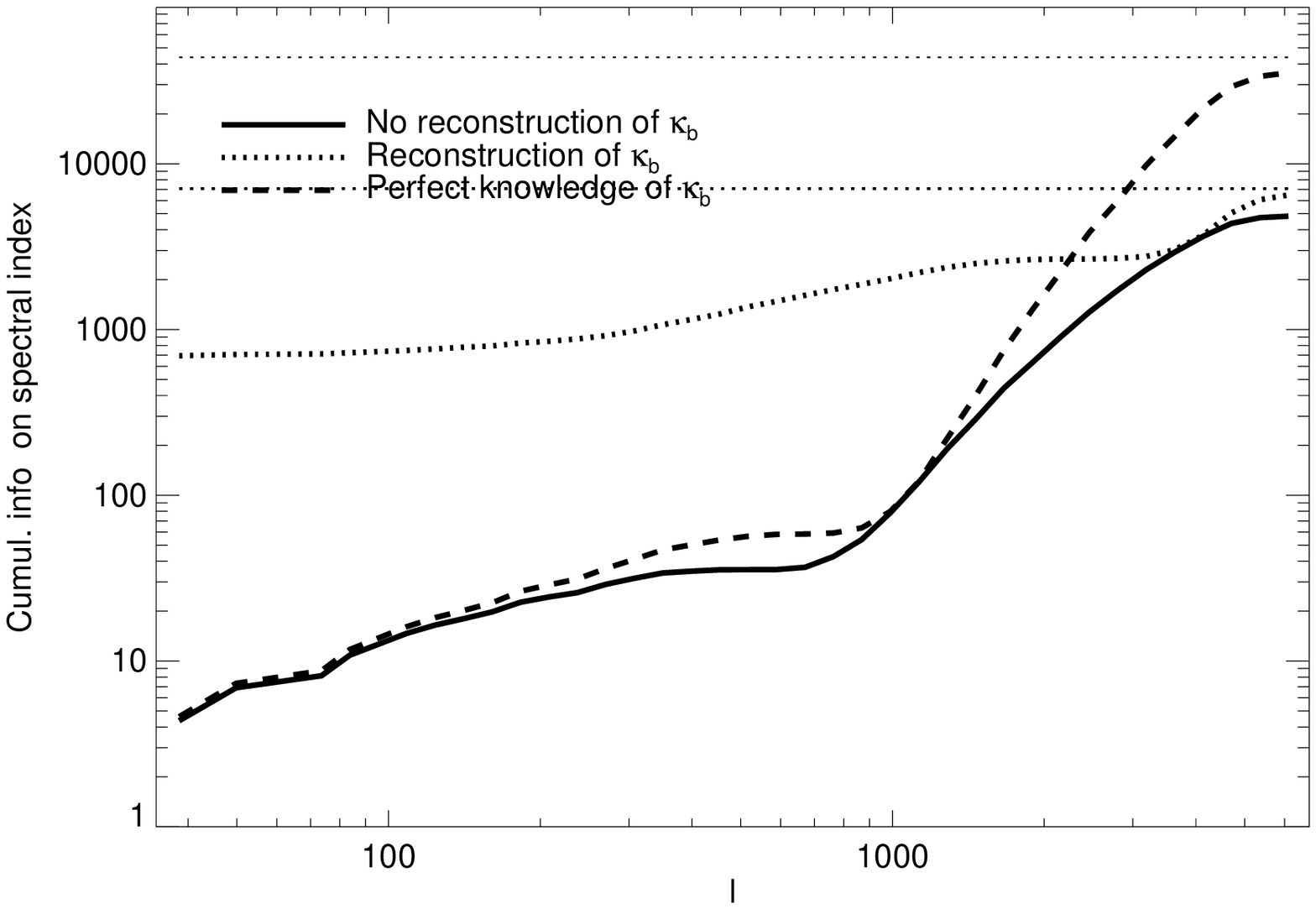}

\caption{The cumulative Fisher information content on $\ln \sigma^2_8$ (left panel) and the spectral index $n_s$ (right panel) of the convergence power spectrum in three different cases, for lognormal convergence field statistics with source redshift $z_s \sim 1$, and a survey volume of $10 \times 10$ deg$^2$. The solid line shows the standard case where no attempt is made at reconstructing the convergence background Fourier mode. The dashed line shows the hypothetical case of a perfect knowledge of this mode, unachievable with shear data, that is then used to recalibrate the spectrum (RC spectrum). The dotted line shows the case where this mode is estimated from the shear field using the statistical dependence of Fourier modes in lognormal fields and used to recalibrate the spectrum. The horizontal thin dotted lines show the total information content of the convergence field (upper line) and that of the convergence field without background mode (lower line) which corresponds to the case of the shear surveys. No shear-based observable can cross the latter line, showing that a large fraction of the total information (here as much as as 90\% for $\sigma^2_8$ and $85\%$ for $n_s$) cannot be captured by the shear fields due to the mass-sheet degeneracy. On the other hand the RC spectrum captures almost all information left accessible. Note that the background mode is always estimated with the full set of Fourier modes. Shape noise is not considered.
\label{infolns28}}
\end{center}
\end{figure}

To understand the extent to which the rescaled power spectrum recaptures information, it is necessary to evaluate its sensitivity to parameters on top of its covariance. The left panel of Fig. \ref{infolns28} shows the cumulative Fisher information of the spectrum
\beq \label{InfoCumul}
F^P_{\alpha}(\le \ell) = \sum_{\ell_1,\ell_2 \le \ell} \frac{\partial \av{\hat P(\ell_1)}}{\partial \alpha } \lb \Sigma_{\le \ell}\rb^{-1}_{\ell_1 \ell_2} \frac{\partial \av{ \hat P(\ell_2)}}{\partial \alpha}. 
\enq
on the linear amplitude $\alpha = \ln \sigma^2_8$ for our three fiducial cases (solid, dotted and dashed). The right panel show the same quantities for the spectral index $\alpha = n_s$.  The derivatives were obtained with the help of finite differences. When simulating the maps with slightly different parameter values, the reconstruction of the background mode still proceeds with the fiducial cosmological parameters as appropriate. The upper horizontal line on each panel shows the total information content $F_{\alpha\alpha}$ of the lognormal map evaluated directly from Eq. \ref{Fln}. The impact of SSC (the difference between solid and dashed) is striking on both panels.  In the case of the RC spectrum, we include to the data vector in Eq. \eqref{InfoCumul} the reconstructed zeroth mode $\hat{\bar \kappa}$ as the first entry. The large scale behavior of the RC curve, the dotted curve, illustrates nicely the importance of the background mode. The RC spectrum unlock small scales information on both panels. It is, in fact, very close to the lower horizontal dotted line, which is the Cramer-Rao bound to which we now turn.

\subsection{Total information content \label{totalinfo}}
To determine whether there is any room for improvement for sophisticated statistical techniques, we now obtain the exact Fisher information content of the PDF describing the observed $\kappa$ map and compare to the information from RC power spectrum.
According to the Cramer-Rao bound, no alternative technique such as higher order statistics or non-linear transforms can outperform the former amount. The exact $d-$variate PDF for $\kappaobs$ is obtained by marginalisation over the unseen $\kappa_b$ as follows
\beq \label{exPDF}
p_\kappaobs(\kappaobs) =  \int_{-\infty}^\infty d\bar\kappa\: p_\kappatrue \lp \kappaobs + \bar \kappa \rp,
\enq
on the surface $\sum_i \kappaobs_i/d = 0$. It has no simpler analytical form in the lognormal model. Nevertheless, we can evaluate its Fisher information content
with direct Monte-Carlo integration, using a large number of simulations such as above. For each simulation we calculate $\partial_\alpha \ln p_\kappa^{\obs}$ that results from Eq. \ref{exPDF}. This is nothing else than the average $\av{\partial_\alpha \ln p_\kappa^{\true}\lp \kappaobs + \bar \kappa\rp}_{\bar \kappa}$ over the almost Gaussian PDF for the zeroth mode such as those shown on Fig. \ref{3PDFs}. We evaluate this average with an accurate Gauss-Legendre quadrature scheme. We then obtain straightforwardly the information $F_{\alpha\alpha} = \av{\lp \partial_\alpha \ln p_\kappa^{\obs}\rp^2}_{\kappaobs}$ averaging over a sufficiently large number of simulations. The result is shown on Fig \ref{infolns28} as the lower horizontal dotted line. It is, in fact, indistinguishable on this figure from the analytical estimation
\beq \label{Fmarg}
F^{\obs}_{\alpha \beta } \simeq F_{\alpha\beta} - \frac{  F_{\alpha \bar\delta} F_{\bar\delta \beta}}{F_{\bar\delta\bar\delta}}
\enq
calculated from the results of section \ref{Modesreconstruction}. This estimation is motivated by marginalization of the background mode, assuming Gaussian statistics for the joint parameter posterior.
 We can conclude that in the lognormal model the mere absence of the single zeroth mode from the data reduces its total information content by factors $\sim 10$ for typical cosmological parameters; this is certainly the most striking result of this paper. Besides, the RC spectrum captures almost optimally the information left available.
\subsection{Full sky tomographic surveys}\label{fullsky}
We found in the last section that a large fraction of the information content of the convergence field is destroyed simply by the absence of the background mode in shear observations. We now show that this effect still holds for full sky surveys and/or in a tomographic setting, with several source redshift bins. In the latter case the cross-correlations can be used to better constrain the background mode corresponding to a given slice. However this introduces as many new background mode parameters as the number of source redshift bins. For simplicity, we consider in the following a log-amplitude parameter with constant log-derivative $\partial_\alpha \ln P_A(\vecell)$, similar to $\sigma_8$, but the same analysis can be performed equally well for any parameter.
\subsubsection{Several source redshifts}
Distributing source galaxies into $N_\textrm{bin}$ redshift bins results in $N_\textrm{bin}$ convergence fields, and as many background modes. The convergence power spectra in the Limber approximation become now 
\beq
P^{nm}_\kappa(\ell) = \int d\chi \frac{g_n(\chi)g_m(\chi)}{\chi^2} P^{3d}\lp \frac \ell \chi,\chi\rp,\quad n,m = 1,\cdots,N_{\textrm{bin}}.
\enq
where $g_n(\chi)$ is the lensing kernel of the corresponding bin. For simplicity we use for each slice the lensing kernel of a single source redshift $\chi_{s,n}$
\beq
g_n(\chi) = \frac 3 2 \lp \frac{H_0}{c}\rp^2 \frac{\chi}{a(\chi)} \lp 1 - \frac{\chi}{\chi_{s,n}} \rp,\textrm{  with total weight  } \kappa_{0,n} = -\int d\chi \: g_n(\chi).
\enq
We then proceed as follows: according to the previous section we use the marginalized Fisher matrix  Eq. \eqref{Fmarg} to evaluate the information content of the observations.  Eq. \eqref{Fmarg}  becomes in the presence of multiple redshift slices
\beq
F^\obs_{\alpha \beta} = F_{\alpha \beta} - \sum_{n,m = 1}^{N_\textrm{bin}} F_{\alpha \bar \delta_n} \lb F^{-1}_{\bar \delta \bar \delta} \rb_{nm} F_{\bar \delta_m \beta}.
\enq
We further define $\epsilon_\alpha = F^{\obs}_{\alpha \alpha} / F_{\alpha \alpha}$, the factor of degradation on the parameter $\alpha$.
\newline
To study the effect of redshift slicing we divide a source redshift interval $0.5$ to $1.5$ equally into a number of bins with $\Nbin$ between $1$ and $10$.  We then evaluate the corresponding information matrices and degradation factors, for a target resolution $\ell_\max = 3258$ well within the multipole range targeted by future weak-lensing surveys (this corresponds to a $d = 128^2$ grid for the $10\times10$ deg$^2$ survey configuration used earlier). To that aim we use Eqs. \eqref{Fdb},\eqref{Fadb} and \eqref{Fln}, where the spectral matrices $P_A$ and $P_\delta$ are given through the circulant embedding relation Eq. \eqref{Pamap}. Fig. \ref{tomofullsky} shows the prediction of $\epsilon$ for a log-amplitude parameter as a function of $\Nbin$, for a periodic $10 \times 10 $ deg$^2$.  and  $20 \times 20 $ deg$^2$, as the dotted and dashed lines respectively. The figure shows that the need to calibrate $N_\textrm{bin}$ parameters totally compensates the additional information provided by the cross-correlation between the bins.
\newline
These predictions still use the flat sky approximation and assume implicitly periodic boundary conditions. We discuss finally the case of full sky coverage.
\subsubsection{Full sky coverage}
\begin{figure}
\begin{center}
\includegraphics[width = 0.45\textwidth]{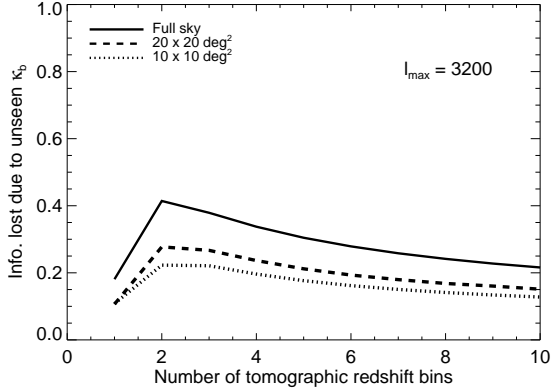}

\caption{\label{tomofullsky} The impact of the mass-sheet degeneracy on cosmological constraints of a full sky cosmic shear survey assuming lognormal field statistics for the convergence field. The solid line shows the prediction of the information degradation on parameters such as $\sigma^2_8$ due solely to the absence of the convergence monopole in shear data. It is shown as a function of the number of source redshift bins between $ z= 0.5$ and $z = 1.5$, and $\ell_\max = 3258$. Shape noise is not considered. The dotted and dashed lines show for comparison the corresponding curves for  smaller surveys of $10 \times 10 $ deg$^2$.  and  $20 \times 20 $ deg$^2$.}
\end{center}
\end{figure}

To approach the limit of full-sky coverage we first discuss the form of the degradation factor. Restricting our attention to a log-amplitude parameter for which $\partial_\alpha \ln P_A(\vecell)$ is a constant and plugging it into the relevant formulae, the degradation is given by (here for $\Nbin = 1$)
\beq \label{degrad}
1-\epsilon =   \frac{ \lp1 + \frac 12  \frac{\sigma^2_A}{\bar n P_A(0)}\rp^2}{\lp \frac 12 + \frac{1}{4} \frac{\sigma^4_A}{\bar n P_A(0)}\rp}    \lb 1 + e^{\sigma^2_A} + \frac{1}{\bar n P_A(0)}+ \frac 1{\bar n} \int \frac{d^2l}{\lp2\pi\rp^2} \lp \frac{P_\delta (\vecell)}{P_A(\vecell)} -1  \rp  \rb^{-1}
\enq
Recall that  $\bar n = d / V$ is the number density of grid points set by the resolution $\ell_\max$ of the observation according to $\bar n  = \ell_\max^2/ 2\pi^2$. It is a useful sanity check that in the Gaussian regime $P_A \rightarrow 0$ we recover $\epsilon = 1$ i.e. no degradation. To that aim, we first note that the prefactor on the right hand side tends to a constant since $\sigma^2_A/P_A(0) $ does so and $\sigma^4_A/P_A(0)$ vanishes. On the other hand the other term is dominated by $1/\bar n P_A(0)$ so that $\epsilon \rightarrow 1$ results. Since $\sigma^2_A$ is fairly small in the cases that interest us the prefactor never plays a key role in this equation. We can see that again $\bar n P_A(0) \sim 1$
is a critical value.   
\newline The key point is that as one might expect all terms entering Eq. \eqref{degrad} are clearly dominated by small scales if the volume is reasonably large, with the only exception of $P_A(0)$.
To obtain the prediction for a full sky survey at a given resolution $\ell_\max$ we can simply evaluate the necessary terms using a smaller volume while replacing $P_A(0) \sim P_\delta(0)$ by the full sky convergence monopole $C^\kappa_{\ell = 0}/\kappa_0^2$. The key quantity becomes the convergence monopole matrix $C_{\ell = 0,nm}$. We evaluate the monopole matrix from its linear theory prediction
\beq
C_{\ell = 0}^{nm} = \frac 2 \pi \int dk\: k^2P^{\textrm{lin.}}_\delta(k,0) W_n(k) W_m(k) \quad \textrm{              with     } \quad
W_n(k) =  \int_0^{\chi_n} d\chi \: g_n(\chi) D(\chi)j_0(k\chi),
\enq
where $D(\chi)$ is the growing mode normalized to unity today, and $j_0(x) = \sin(x)/x$ the zero-th spherical Bessel function.
We then obtained in the way exposed above the factor of degradation for each bin configuration investigated above. Fig. \ref{tomofullsky} shows as the solid line the prediction of $\epsilon$ as a function of $\Nbin$. Table \ref{tab1} shows as the lower left corner the value of the monopole matrix $\bar n C^{nm}_{\delta,\ell = 0}$ for the source redshifts value given in the first line. Shown in the upper right corner are the corresponding values for the $10\times10$ deg$^2$ survey. The monopole matrix is somewhat reduced in the case of the full sky survey, but  is still always similar to or greater than unity, explaining the trend seen on Fig. \ref{tomofullsky}. We can safely conclude that neither a tomographic analysis nor the observation of the full sky changes our main points. 
\setcounter{table}{0}
\begin{table*}
\begin{minipage}{180mm}

\caption{The lower left corner shows the monopole matrix $\bar n C_{\delta,\vecell= 0}$, for a full sky survey with source redshifts as given in the first line and target resolution $\ell_\max \sim 3000$. The upper right corner shows for comparison the corresponding matrix for a $10 \times 10 $ deg$^2$. Values similar or greater than unity, which we conclude is always the case in a $\Lambda$CDM universe even for full sky surveys indicate that the mass-sheet degeneracy degrades substantially achievable constraints on typical cosmological parameters following the equations given in the text. $\bar n $ is $\ell_\max^2/2\pi^2$.}
\label{tab1}
\begin{tabular}{lccccc}

Source redshifts $z_s$ & 0.50& 0.75& 1.00& 1.25& 1.50\\  
\hline

&  23 / 28 & 17 & 11 & 7.8 & 6.0   \\ 
&  12 & 6.1 / 11 & 7.3 & 5.4 & 4.2   \\ 
Monopole matrix $\bar n C_{\ell = 0}$&  7.1 & 3.8 & 2.4 / 5.3 & 4.0 & 3.1   \\ 
for full sky vs $10 \times 10$ deg$^2$ survey&  4.9 & 2.6 & 1.7 & 1.2 / 3.1 & 2.4  \\ 
&  3.7 & 2.0 & 1.3 & 0.9 & 0.7 / 2.0   \\ 

 \hline
 Minimal convergence $\kappa_0$ & -0.019& -0.040& -0.067& -0.097& -0.130\\ \hline
\end{tabular}
\end{minipage}
\end{table*}

\section{Conclusions and discussion} \label{conclusion}
We estimated analytically the impact of super-survey modes on the information content of cosmic convergence fields and tested our approximations with simulations. We worked within the assumption of lognormal field statistics and derived the information content of the field on the background mode, and if considered as a nuisance parameter,  its degeneracy with cosmological parameters. We paid particular attention to the impact of the absence of the zero mode of the convergence determined from noise-free shear fields due to the mass-sheet degeneracy.
\newline
\indent 
Super-survey modes can be treated as parameters similarly to cosmological parameters and marginalized over. Following this idea, our principal result is that the absence of a single mode, the monopole, from the shear fields has drastic consequences on its information content: perhaps surprisingly,  $80\% -90\%$ of the \textit{total} information of the field vanishes in an unrecoverable fashion for typical cases.  This translates into a degradation by a factor of $2-3$ of achievable constraints on cosmological parameters. 
\newline
\indent 
Standard convergence power spectra capture even less information. We showed that the statistical dependence of non-Gausssian Fourier modes allow the reconstruction of the unobserved background mode from the observed small scales modes. Knowledge of this mode allows recalibration of the power spectrum, in analogy to the case of the galaxy power spectrum \citep{DePutterEtal2012}, through the introduction of a renormalized convergence (RC) map. The recalibration we propose diagonalizes the covariance matrix and at the same time moderates the impact of SSC.  In the configurations and methods investigated in this paper, we found that the RC spectrum can outperfom the standard power spectrum by a factor of two.
\newline
\indent
During completion of this work we became aware of the recent preprint \cite{LiEtal2014b}, that investigates joint reconstruction of the background mode and cosmological parameters in $N$-body simulations from the matter power spectrum only. Showing comparable factors of degradation, our results are fully consistent with theirs, and show how this is a natural consequence of the dynamics of the information within the full field, Eqs. \eqref{Fdb},\eqref{Fadb} and \eqref{Fln}. 
One claim of \cite{LiEtal2014b} that might seem at first incompatible to our results is that the spectrum defined with respect to the local density suffers more from degeneracies to the background mode than the spectrum defined with respect to the global density. To avoid confusion it should be noted, however, that in the local case \cite{LiEtal2014b} reconstructs all parameters (inclusive the background mode) from the local spectrum. Our method first reconstructs the background mode and making profit of that knowledge we turn the 'global' power spectrum into a local (RC) power spectrum, improving its sensitivity to traditional cosmological parameters. Clearly we do not use it to infer the background mode itself. It is intuitively obvious that in the sole local spectrum is not a suitable observable in order to constrain the background mode (as \cite{LiEtal2014b} in effect show), as it is by design less sensitive to it \citep{DePutterEtal2012} than the global spectrum.
\newline
\indent
While our simple modification of the power spectrum achieves significant gain for cosmological constraints, the question is whether more advanced statistical techniques can result in further improvement. We checked carefully that the marginalisation procedure is justified, by successful comparison to the true total information content of the observed convergence maps. Since the latter gives the fundamental Cramer-Rao bound, we find that no other technique, such as higher-order statistics or optimal non-linear transforms of the convergence,\citep{SeoEtal2011,JoachimiEtal2011,Carron2012}, not even the sufficient statistics \citep{CarronSzapudi2013, CarronSzapudi2014} adapted to shear fields can yield further significant improvement over our RC spectra, even for full sky surveys. Only additional data providing information on the convergence monopole provide an alternative route to alleviate the information loss from SSC effect. Such techniques might be built for instance on combining properly shear and amplification measurements. The study and construction of such statistics is left for future work.
\newline
\indent
In order to obtain analytical results, and to make use of fast simulation techniques, we
assumed lognormal statistics with periodic boundary conditions and did not consider shape noise. Our conclusions might need to be modified slightly for realistic surveys, a task left for future work. In particular, we focused on the zero mode, however, edge effects due to survey geometry cause leakage of low-$\ell$ modes introducing off-diagonal elements in the covariance matrix even in the case of a Gaussian process. Moreover, shape noise complicates the lognormal likelihood we used to reconstruct the zero mode, and this could affect quantitatively the effectiveness of the RC power spectrum. Finally, the lognormal model is only an approximation, in realistic cases often super-lognormal tails are observed \citep{DasOstriker2006,TakahashiEtal2011}, thus there might still be some room for advanced statistical techniques, such as sufficient statistics to squeeze all cosmological information from the data. Nevertheless, lognormal statistics captures remarkably well the information content and covariances of the dark matter field, as it has been demonstrated in a similar context \citep{CarronNeyrinck2012,CarronEtal2014} earlier, giving invaluable analytical insights on the inherently non-Gaussian properties of the dark matter field. We are confident that the above model is realistic enough that they shed light on the surprisingly large information loss due to the missing information on large scale modes due to the mass sheet degeneracy, the principal result of this investigation.


\section*{Acknowledgments}
The authors acknowledge NASA grants NNX12AF83G and NNX10AD53G for support, thank Melody Wolk for many useful discussions as well as the anonymous referee for useful comments improving the presentation of the results of this paper.
\appendix
\section{Information on background mode and its correlation to cosmological parameters} \label{AppendixInfo}
We derive in this section the fundamental Eqs \eqref{Fdb} and \eqref{Fadb}  starting from the expressions for a generic field PDF, Eqs. \eqref{Fdeltab} and \eqref{Fdeltabparam}. The case of a zero ensemble mean Gaussian field with two-point function $\xi_\delta(x_i,x_j)$ (a $d \times d$ matrix $\xi_{\delta,ij}$) is of  a useful warm up exercise and straightforward. From $\ln p_\delta = -\frac 12 \delta^t \xi_\delta^{-1} \cdot \delta +\cst.$ follows
\beq
-\frac{\partial \ln p_\delta}{\partial \bar \delta}  =  \vecun^t \xi_{\delta}^{-1} \cdot \delta \quad \textrm{and} \quad -\frac{\partial^2 \ln p_\delta}{\partial \bar \delta^2}   = \vecun^t \xi_{\delta}^{-1} \cdot \vecun.
\enq
We have introduced the vector $\vecun = (1,\cdots,1)^t$. Taking the average results in
\beq \label{eqA2}
F_{\alpha \bar \delta} = 0, \quad F_{\bar \delta \bar \delta} = \vecun^t \xi_{\delta}^{-1} \cdot \vecun
\enq
On a regular grid with periodic boundary conditions (or full sky coverage), the last sum results in
\beq
 F_{\bar \delta \bar \delta} = \vecun^t \xi_{\delta}^{-1} \cdot \vecun =\lp \frac V {P_\delta(0)} \rp , \textrm{  where  } \av{\tilde \delta_\veck \tilde \delta^{*}_{\veck'}} = \delta_{\veck\veck^{'}}VP_\delta(\veck).
\enq
Due to the statistical independence of modes in the Gaussian field, the information on the background mode is simply its inverse variance in the volume, as expected.
\subsection{Background mode in the lognormal field}
The lognormal field is defined through $\ln p_\delta = -\frac 12 (A -\bar A)^t \xi_A^{-1} \cdot (A -\bar A) - \vecun \cdot A +\cst,$ where $A = \ln (1 + \delta)$ is the Gaussian log-density field. For the purpose of this section $\bar A = -\sigma^2_A/2$ denotes the ensemble mean of $A$ and no spatial average.The second term in $\ln p_\delta$ is the Jacobian of the transformation from $\delta$ to $A$. Clearly, $\partial A_i /\partial \delta_j = \delta_{ij}/(1 + \delta_j) = \delta_{ij}e^{-A_j}$. It follows
\beq
-\frac{\partial \ln p_\delta}{\partial \bar \delta} = (A - \bar A)^t \xi_A^{-1}\cdot e^{-A} + \vecun \cdot e^{-A}.
\enq
and
\beq
-\frac{\partial^2 \ln p_\delta}{\partial \bar \delta^2} = e^{-A,t}  \xi_A^{-1}\cdot e^{-A} - (A -\bar A)^t\xi_A^{-1}\cdot e^{-2A}  - \vecun \cdot e^{-2A}.
\enq
Also,
\beq
-\frac{\partial^2 \ln p_\delta}{\partial \bar \delta \partial \alpha} = -(A  -\bar A)^t\xi^{-1}_A\frac{\partial \xi_A}{\partial \alpha}\xi_A^{-1}\cdot e^{-A} - \frac{\partial \bar A}{\partial \alpha} \:\vecun^t \xi_A^{-1}\cdot e^{-A}.
\enq
To obtain the information we need to obtain the average of these expressions with respect to the Gaussian PDF for $A$.
It is useful to derive an intermediate result as follows. Define the vector $v = \xi_A^{-1} (A - \bar A)$ and the density map $\rho = \exp(A)$. All needed quantities can be written in terms of averages of the form
\beq
\av{\rho^{\vecm}v^\vecn} = \left. \frac{\partial^{|\vecn|}}{\partial s^\vecn} \right|_{s = 0}\av{e^{\vecm \cdot A + s^t\xi_A^{-1}\cdot(A -\bar A) }}
\enq
for multiindices $\vecn = (n_1,\cdots,n_d)$, $\vecm = (m_1,\cdots,m_d)$. The expectation value is now a standard Gaussian integral, that results in
\beq
\av{\rho^\vecm v^\vecn} = \av{\rho^\vecm} \left.\frac{\partial^{|\vecn|}}{\partial s^\vecn} \right|_{s = 0} e^{\frac 12 s^t\xi_A^{-1}\cdot s + s\cdot \vecm} \quad \textrm{with} \quad \av{\rho^\vecm} = \exp\lp \bar A\: \vecun\cdot \vecm + \frac 12 \vecm^t \xi_A \cdot \vecm \rp.
\enq
The quantities we need follow directly
\beq
\av{v_i \frac{1}{\rho_j}} = -\delta_{ij}\av{\frac 1 {\rho_j}}, \quad \av{v_i \frac{1}{\rho^2_j}} = -2\delta_{ij}\av{\frac 1 {\rho_j^2}}
 \quad \textrm{and} \quad  \av{\frac{1}{\rho_i}} = e^{\sigma^2_A}, \quad \av{\frac{1}{\rho_i\rho_j}} = e^{2\sigma^2_A +\xi_{A,ij}} = e^{2\sigma^2_A}\lp 1 + \xi_{\delta,ij}\rp.
\enq
Putting all this together gives in general
\beq
F_{\bar \delta \bar \delta} = e^{2\sigma^2_A} \lb d \:e^{\sigma^2_A} + \vecun^t \xi_A^{-1}\cdot\vecun + \Tr \:\xi_A^{-1}\xi_\delta \rb \quad \textrm{and} \quad F_{\alpha \bar \delta} = e^{\sigma^2_A}\lb \frac 12 \frac{\partial \sigma^2_A}{\partial \alpha} \vecun^t \xi_A^{-1} \vecun +  \Tr \: \xi_A^{-1}\frac{\partial \xi_A}{\partial \alpha} \rb.
\enq
For a periodic volume we can simplify further to
\beq
F_{\bar \delta \bar \delta} = e^{2\sigma^2_A} \lb d \:e^{\sigma^2_A} + \lp \frac V {P_A(0)}  \rp+ \sum_\veck \frac{P_\delta(\veck)}{P_A(\veck)}\rb \quad \textrm{and} \quad F_{\alpha \bar \delta} = e^{\sigma^2_A}\lb \frac12\frac{\partial \bar \sigma^2_A}{\partial \alpha} \lp \frac{V}{P_A(0)}\rp +  \sum_\veck \frac{\partial \ln P_A(\veck)}{\partial \alpha} \rb.
\enq
The case of a tomographic configuration with several jointly lognormal fields is not more difficult. The only differences being the covariance matrices $\xi_{ij}$ at fixed $x_i,x_j$ are now $\Nbin$ dimensional matrices as discussed in the text. This leads straightforwardly to the expressions in the text.

\subsection{Including $\kappa_0$ dependence}
In the lognormal model of the convergence field the density fluctuation $\delta =  \kappa/|\kappa_0|$ carries a cosmological parameter dependence through the geometrical factors entering $|\kappa_0|$. In this paper we investigated $\sigma_8$ and $n_s$ for which there is no such dependence. For completeness we present here the above results in the most general case.
\newline
The presence of $\kappa_0$ introduces additional terms according to $\partial_\alpha A = \lp e^{-A} - 1\rp \partial_\alpha \ln \kappa_0^2/2$. The resulting terms are thus of precisely the same type than for the background mode calculation, and all the necessary results are present above. We simply state the final results. The Fisher matrix on cosmological parameters can now be decomposed as the sum of three positive terms as follows
\beq
\begin{split}
F_{\alpha\beta} &=  \frac V {P_A(0)}\lp \frac{1}{2} \frac{\partial \sigma^2_A}{\partial \alpha} + \frac 12 \frac{\partial \ln \kappa_0^2}{\partial \alpha}\sigma^2_\delta \rp \lp  \alpha \leftrightarrow \beta \rp + \frac 12 \sum_\veck \lb \lp \frac{\partial \ln P_A(\veck)}{\partial \alpha} + e^{\sigma^2_A} \frac{\partial \ln \kappa_0^2}{\partial \alpha} \rp \lp \alpha \leftrightarrow \beta \rp \rb \\
& + \frac {e^{2\sigma^2_A}}4 \frac{\partial \ln \kappa_0^2}{\partial \alpha}\frac{\partial \ln \kappa_0^2}{\partial \beta} \lp d(e^{\sigma^2_A}  - 1) + \sum_\veck \lp \frac{P_\delta(\veck)}{P_A(\veck)} -1\rp  \rp.
\end{split}
\enq
Likewise, the degeneracy between background modes and parameters breaks down into
\beq
F_{\alpha \bar \delta} = e^{\sigma^2_A}\lb \frac 12 \frac{\partial \sigma^2_A}{\partial \alpha}  \lp \frac V {P_A(0)} \rp + \sum_\veck \frac{\partial \ln P_A(\veck)}{\partial \alpha} \rb  + \frac 12 \frac{\partial \ln \kappa_0^2}{\partial \alpha} \lb \lp \frac V {P_A(0)} \rp \lp \sigma^4_\delta + \sigma^2_\delta \rp + de^{3\sigma^2_A} + e^{2\sigma^2_A} \sum_\veck \frac{P_\delta(\veck)}{P_A(\veck)} \rb.
\enq 
\bibliographystyle{mn2e}
\bibliography{bib}

\end{document}